\documentstyle[12pt,procsla]{article}
  
  \catcode`\@=11
  \long\def\@makefntext#1{
  \protect\noindent \hbox to 3.2pt {\hskip-.9pt  
  $^{{\ninerm\@thefnmark}}$\hfil}#1\hfill}              %CAN BE USED 
  
  \def\@makefnmark{\hbox to 0pt{$^{\@thefnmark}$\hss}}  %ORIGINAL 
        
  \def\ps@myheadings{\let\@mkboth\@gobbletwo
  \def\@oddhead{\hbox{}
  \rightmark\hfil\ninerm\thepage}   
  \def\@oddfoot{}\def\@evenhead{\ninerm\thepage\hfil
  \leftmark\hbox{}}\def\@evenfoot{}
  \def\sectionmark##1{}\def\subsectionmark##1{}}
\def\gappeq{\mathrel{\rlap {\raise.5ex\hbox{$>$}} {\lower.5ex\hbox{$\sim$}}}}
\def\lappeq{\mathrel{\rlap{\raise.5ex\hbox{$<$}} {\lower.5ex\hbox{$\sim$}}}}
\def\beq{\begin{equation}} 
\def\eeq{\end{equation}} 
\def\bea{\begin{eqnarray}}
\def\eea{\end{eqnarray}}
\def\bq{\begin{quote}} 
\def\eq{\end{quote}}
\def\dd{\displaystyle}
\def\ov{\overline}
\def\um{1/2}
\def\sqr{\sqrt{2}}
\def\sq{1/\sqrt{2}}
\def\eV{{\rm eV}}
\def\KeV{{\rm KeV}}
\def\MeV{{\rm MeV}}
\def\GeV{{\rm GeV}}
\def\sq{1/\sqrt{2}}

\begin{document}
  \begin{flushright} {CERN-TH/99-129 \\ DFPD-99/TH/21} \end{flushright} 
\vspace*{15mm}
    \centerline{\normalsize\bf NEUTRINO MASSES AND MIXINGS: A THEORETICAL PERSPECTIVE}
  \baselineskip=16pt
 % \centerline{\normalsize\bf MANUSCRIPT BY COMPUTER}
 % \centerline{\footnotesize\sf (For subsequent $\sim$ 15\% photoreduction)
 % \footnote{The \LaTeX\ source
 % file for this document may be used as a template for your
 % article.}
 % }  
  
  %\vfill
  \vspace*{0.6cm}
  \centerline{\footnotesize GUIDO ALTARELLI}
  \baselineskip=13pt
  \centerline{\footnotesize\it Theoretical Physics Division, CERN, CH - 1211 Geneva 23, Switzerland}
  \baselineskip=12pt
  \centerline{\footnotesize\it and}
  \baselineskip=12pt
  \centerline{\footnotesize\it Universit\`a di Roma Tre, Rome, Italy}
  \centerline{\footnotesize E-mail: guido.altarelli@cern.ch}
  \vspace*{0.3cm}
  \centerline{\footnotesize and}
  \vspace*{0.3cm}
  \centerline{\footnotesize FERRUCCIO FERUGLIO}
  \baselineskip=13pt
  \centerline{\footnotesize\it Universit\`a di Padova and I.N.F.N., Sezione di Padova, Padua, Italy}
  \centerline{\footnotesize E-mail: feruglio@pd.infn.it}
  
  %\vfill
  \vspace*{0.9cm}
  \abstracts{We briefly review the recent activity on neutrino masses and mixings which was prompted by the
  confirmation of neutrino oscillations by the Superkamiokande experiment.}

\vspace*{1.8cm}
\noindent
PACS: 11.30.Hv, 12.10.-g, 12.15.Ff, 14.60.Pq
\vspace*{0.2cm}

\noindent
Keywords: Solar and Atmospheric Neutrinos, Beyond the Standard Model, Neutrino Physics, Grand Unified Theories
   
  %\vspace*{0.6cm}
  \normalsize\baselineskip=15pt
  \setcounter{footnote}{0}
  \renewcommand{\thefootnote}{\alph{footnote}}
\section{Introduction}

It is for us a great pleasure to contribute to the celebration of Lev Okun anniversary with this article. Considering the
continuous interest of Lev on neutrinos we thought that this subject is particularly appropriate to the occasion.

Recent data from Superkamiokande \cite{SK} have provided a more solid experimental basis for neutrino oscillations as an
explanation of the atmospheric neutrino anomaly. In addition the solar neutrino deficit, observed by several experiments
\cite{solexp}, is also probably an indication of a different sort of neutrino oscillations. Results from the laboratory
experiment by the LSND collaboration \cite{LSND} can also be considered as a possible indication of yet another type of
neutrino oscillation.  Neutrino oscillations 
\footnote{Alternative explanations such as neutrino decay and violations of the equivalence principle appear to be disfavoured 
by the present data \cite{altern}.}~
imply neutrino masses. The extreme smallness of neutrino masses in comparison
with quark and charged lepton masses indicates a different nature of neutrino masses, linked to lepton number violation and the
Majorana nature of neutrinos. Thus neutrino masses provide a window on the very large energy scale where lepton number is
violated and on Grand Unified Theories (GUTs) \cite{ross}. The new experimental evidence on neutrino masses could also give an
important feedback on the problem of quark and charged lepton masses, as all these masses are possibly related in GUTs. In
particular the observation of a nearly maximal mixing angle for $\nu_{\mu}\rightarrow \nu_{\tau}$ is particularly interesting.
Perhaps also solar neutrinos may occur with large mixing angle. At present solar neutrino mixings can be either large or very
small, depending on which particular solution will eventually be established by the data. Large mixings are very interesting
because a first guess was in favour of small mixings in the neutrino sector in analogy to what is observed for quarks. If
confirmed, single or double maximal mixings can provide an important hint on the mechanisms that generate neutrino masses. The
purpose of this article is to provide a concise review of the implication of neutrino masses and mixings on our picture of
particle physics. We will not review in detail the status of the data but  rather concentrate on their conceptual impact. 

The experimental status of neutrino oscillations is still very preliminary. While the evidence for the existence of neutrino
oscillations from solar and atmospheric neutrino data is rather convincing by now, the values of the mass squared differences
$\Delta m^2$ and mixing angles are not firmly established. For the observed $\nu_e$ suppression of solar neutrinos,  for
example, three possible solutions are still possible \cite{solar}. Two are based on the MSW mechanism \cite{MSW}, one with
small (MSW-SA:
$\sin^2{2\theta_{sun}}\sim 5.5~10^{-3}$) and one with large mixing angle (MSW-LA:  $\sin^2{2\theta_{sun}}\gappeq 0.2$). The
third solution is in terms of vacuum oscillations (VO) with large mixing angle (VO:
$\sin^2{2\theta_{sun}}\sim 0.75$). However, it is important to keep in mind that the $\Delta m^2$ values of the above
solutions are determined by the experimental result that the suppression is energy dependent. This is obtained by comparing
experiments with different thresholds. The Cl experiment shows a suppression larger than by a factor of 2, which is what is
shown by Ga and water experiments \cite{solexp}. If the Cl indication is disregarded, then new energy independent solutions
would emerge, with large $\Delta m^2$ and maximal mixing. For example, good fit of all data, leaving those on Cl aside, can be
obtained with
$\Delta m^2$ as large as $\Delta m^2\lappeq 10^{-3} \eV^2$ \cite{stru,har}. For atmospheric neutrinos the preferred value of
$\Delta m^2$, in the range $10^{-3}-10^{-2}~\eV^2$, is still affected by experimental uncertainties and could sizeably
drift in one sense or the other, but the fact that the mixing angle is large appears established ($\sin^2{2\theta_{atm}}\gappeq
0.9 ~{\rm at}~ 90\%~ {\rm C.L.}$)
\cite{fogli,hall,WIN99}. Another issue which is still open is the claim by the LSND collaboration of an additional  signal of
neutrino oscillations in a terrestrial experiment \cite{LSND}. This claim was not so-far supported by a second recent experiment,
Karmen \cite{Karmen}, but the issue is far from being closed. 

Given the present experimental uncertainties the theorist has to
make some assumptions on how the data will finally look like in the future. Here we tentatively assume that the LSND evidence
will disappear. If so then we only  have two oscillations frequencies, which can be given in terms of the three known species
of light neutrinos without additional sterile kinds (i.e. without weak interactions, so that they are not excluded by LEP). We
then take for granted that the frequency of atmospheric neutrino oscillations will remain well separated from the solar
neutrino frequency, even for the MSW solutions. The present best values are \cite{solar,fogli,hall,WIN99,ue3} 
$(\Delta m^2)_{atm}\sim 3.5~10^{-3}~\eV^2$ and $(\Delta m^2)_{MSW-SA}\sim 5~10^{-6}~\eV^2$ or
$(\Delta m^2)_{VO}\sim 10^{-10}~\eV^2$. We also assume that the electron neutrino does not participate in the atmospheric
oscillations, which (in absence of sterile neutrinos) are interpreted as nearly maximal
$\nu_{\mu}\rightarrow\nu_{\tau}$ oscillations as indicated by the Superkamiokande \cite{SK} and Chooz \cite{Chooz} data.
However the data do not exclude a non-vanishing $U_{e3}$ element. In the Superkamiokande allowed region the bound by Chooz
\cite{Chooz} amounts to  $|U_{e3}|\lappeq 0.2$ \cite{fogli,hall,ue3}.

\section{Direct Limits on Neutrino Masses}

Neutrino oscillations are due to a misalignment between the flavour basis, $\nu'\equiv(\nu_e,\nu_{\mu},\nu_{\tau})$, where
$\nu_e$ is the partner of the mass and flavour eigenstate $e^-$ in a left-handed weak isospin $SU(2)$ doublet (similarly for 
$\nu_{\mu}$ and $\nu_{\tau})$) and the neutrino mass eigenstates $\nu\equiv(\nu_1, \nu_2,\nu_3)$: 
\beq
\vert\nu'\rangle =U\vert\nu\rangle,\label{U}
\eeq where $U$ is the 3 by 3 mixing matrix \cite{MNS}. 
Thus, in presence of mixing, neutrinos cannot be all massless and actually the
presence of two different oscillation frequencies implies at least two different non zero masses. Neutrino
oscillations are practically only sensitive to differences $\Delta m^2$ so that the absolute scale of squared masses is not
fixed by the observed frequencies. But the existing direct bounds on neutrino masses, together with the observed frequencies,
imply that all neutrino masses are by far smaller than any quark or lepton masses. In fact the following direct bounds hold:
$m_{\nu_e}\lappeq \sim 5~\eV$, $m_{\nu_{\mu}}\lappeq 170~ \KeV$ and $m_{\nu_{\tau}}\lappeq 18~ \MeV$ \cite{PDG}. Since the
observed
$\Delta m^2$ indicate mass splittings much smaller than that, the limit on $m_{\nu_e}$ is actually a limit on all neutrino
masses. Moreover from cosmology we know \cite{tur} that the sum of masses of (practically) stable neutrinos cannot exceed a few
eV, say
$\sum m_{\nu_i}\lappeq 6~\eV$, corresponding to a fraction of the critical density for neutrino hot dark matter
$\Omega_{\nu} h^2\lappeq 0.06$ (the present value of the reduced Hubble constant $h$ being around 0.7). In conclusion, the
heaviest light neutrinos that are allowed are three nearly degenerate neutrinos of mass around or somewhat below $2~\eV$. In
this case neutrinos would be of cosmological relevance as hot dark matter and contribute a relevant fraction of the critical
density.   But at present there is no compelling experimental evidence for the necessity of hot dark matter \cite{tur,kra}.  As
a consequence, neutrino masses can possibly be much smaller than that. In fact, for widely split neutrino masses the heaviest
neutrino would have a mass around
$\sim 0.06~\eV$ as implied by the atmospheric neutrino frequency. 

An additional important constraint on neutrino masses, which will be relevant in the following, is obtained from the non
observation of neutrino-less double beta decay. This is an upper limit on the $\nu_e$ Majorana mass, or equivalently, on
$m^{ee}_{\nu}=\sum_i U^2_{ei} m_i$, which is at present quoted to be $m^{ee}_{\nu}\lappeq 0.2~\eV$ \cite{dbeta}.  

\section{Neutrino Masses and Lepton Number Violation}

Neutrino oscillations imply neutrino masses which in turn demand either the existence of right-handed neutrinos (Dirac masses)
or lepton number  (L) violation (Majorana masses) or both. Given that neutrino masses are extremely small, it is really
difficult from the theory point of view to avoid the conclusion that L must be violated. In fact, it is only in terms of lepton
number violation that the smallness of neutrino masses can be explained as inversely proportional to the very large scale where
L is violated, of order $M_{GUT}$ or even $M_{Planck}$.

Recall that an ordinary Dirac mass term is of the form $m_{RL}=\ov{\nu_R} m_D \nu_L$ plus its hermitian conjugate $m_{LR}$,
where
$\nu_{R,L}=1/2 (1\pm \gamma_5) \nu$ and $\nu$ is the neutrino field. The field $\nu_L$ annihilates a left-handed neutrino and
creates a right-handed antineutrino. Correspondingly the field $\ov{\nu_L}$ creates a left-handed neutrino and annihilates a
right-handed antineutrino. Now left-handed neutrinos $\nu_L$ and right-handed antineutrinos $(\bar\nu)_R$ are indeed the only
observed neutrino states. Thus in principle one could assume that right-handed neutrinos and left-handed antineutrinos do not
exist at all. The field $\nu_L$ and its conjugate $\ov{\nu_L}$ would then suffice and there would be no $\nu_R$ and 
$\ov{\nu_R}$ fields in the theory. In the SM lagrangian density only the term $i\ov{\nu_L} D\llap{$/$}\nu_L$ appears, where
$D_{\mu}$ is the $SU(2) \bigotimes U(1)$ gauge covariant derivative. Clearly if no
$\nu_R$ is allowed there is no possible Dirac mass for a neutrino. 

But if L is violated, there is no conserved quantum number that really makes neutrinos and antineutrinos different and a new
type of mass term is possible. For a massive neutrino, the positive helicity state that Lorentz invariance demands to be
associated with the state
$\nu_L$ of given momentum can be $(\bar\nu)_R$ (for a charged particle with mass, say an $e^-_L$, one can go to the rest frame,
rotate the spin by
$180^o$ and boost again to the original momentum to obtain $e^-_R$). If neutrinos and antineutrinos are not really distinct
particles, both Lorentz and TCP invariance are satisfied by just allowing
$\nu_L$ and $(\bar\nu)_R$ (TCP changes one into the other at fixed momentum). For a massive charged particle of spin 1/2 one
needs four states, while only two are enough for an intrinsically neutral particle. If L is violated we can have a Majorana
mass term
$m_{LL}=\ov{(\nu^c)_R} m \nu_L=\nu^T_LC m \nu_L$ where $(\nu^c)_R=C \ov{\nu_L}^T$ and $C$ is the 4 by 4 matrix in Dirac space
that implements charge conjugation (the field $(\nu^c)_R$ annihilates a $(\bar\nu)_R$ exactly as the field
$\ov{\nu_L}^T$ does, the transposition only indicating that we want it as a column vector). Clearly, the Majorana mass term
$m_{LL}$ violates L by two units. Also, since in the SM $\nu_L$ is a weak isospin doublet, $m_{LL}$ transforms as a component
of an isospin triplet. In the following, as we are only interested in flavour indices and not in Dirac indices, we will simply
denote $m_{LL}$ by $m_{LL}=\nu^T_L m \nu_L$, omitting the Dirac matrix $C$. Note that if L is violated and
$\nu_R$ also exists, then a second type of Majorana mass is also possible which is $m_{RR}=\nu^T_R m \nu_R$, where we again
omitted $C$. Clearly also $m_{RR}$ violates L by two units, but, since $\nu_R$ is a gauge singlet, $m_{RR}$ is invariant under
the SM gauge group. In conclusion, if $\nu_R$ does not exist, we can only have a Majorana mass $m_{LL}$ if L is violated. If
$\nu_R$  exists and L is violated, we can have both Dirac $m_{LR}$ and Majorana masses $m_{LL}$ and $m_{RR}$.

Imagine that one wanted to give masses to neutrinos and, at the same time, avoid the conclusion that lepton number is violated.
Then he/she must assume that $\nu_R$ exists and that neutrinos acquire Dirac masses through the usual Higgs mechanism as quark
and leptons do. Technically this is possible. But there are two arguments against this possibility. The first argument is that
neutrino masses are extremely small so that the corresponding Yukawa couplings would be enormously smaller than those of any
other fermion. Note that within each generation the spread of masses is by no more than a factor $10^{1-2}$. But the spread
between the $t$ quark and the heaviest neutrino would exceed a factor of $10^{11}$. A second argument arises from the fact that
once we introduce $\nu_R$ in the theory, then the L violating term
$m_{RR}=\nu^T_R m \nu_R$ is allowed in the lagrangian density by the gauge symmetry. In the minimal SM, i.e. without
$\nu_R$,  we understand L and B conservation as accidental global symmetries that hold because there is no operator term of
dimension
$\leq4$ that violates B and L but respects the gauge symmetry. For example, the transition $u+u \rightarrow e^++\bar d$ is
allowed by colour, weak isospin and hypercharge gauge symmetries, but corresponds to a four-fermion operator of dimension 6:
$O_6=(\lambda/M^2) (e^Td)(u^Tu)$. This term is suppressed by the dimensional factor $1/M^2$. In the assumption that the SM
extended by Supersymmetry is an effective low energy theory which is valid up to the GUT scale, as suggested by the
compatibility of observed low energy gauge couplings with the notion of unification at $M_{GUT}\sim10^{16} ~\GeV$, the large
mass $M$ is identified with
$M_{GUT}$. In fact the factor $1/M^2_{GUT}$ can be obtained from the propagator of a superheavy intermediate gauge boson with
the right quantum numbers (e.g. those of an SU(5) generator). In supersymmetric models with R invariance the status of B and L
conservation as an accidental symmetry is maintained. In the presence of $\nu_R$, the dimension 3 operator corresponding to
$m_{RR}$ is gauge symmetric but violates L. By dimensions we expect a mass factor in front of this operator in the lagrangian
density, and in the absence of a protective symmetry, we expect it of the order of the cut-off, i.e. of order $M_{GUT}$ or
larger. Thus, L number violation is naturally induced by the presence of $\nu_R$, unless we enforce it by hand.

\section{L Violation Explains the Smallness of Neutrino Masses: the See-Saw Mechanism} 

Once we accept L violation we gain an elegant explanation for the smallness of neutrino masses as they turn out to be inversely
proportional to the large scale where lepton number is violated. If L is not conserved, even in the absence of
$\nu_R$, Majorana masses can be generated for neutrinos by dimension five operators of the form 
\beq O_5=L^T_i\lambda_{ij}L_j\phi\phi/M\label{O5}
\eeq with $\phi$ being the ordinary Higgs doublet, $\lambda$ a matrix in flavour space and $M$ a large scale of mass, of order
$M_{GUT}$ or $M_{Planck}$. Neutrino masses generated by $O_5$ are of the order
$m_{\nu}\sim v^2/M$ for $\lambda_{ij}\sim {\rm O}(1)$, where $v\sim {\rm O}(100~\GeV)$ is the vacuum expectation value of the
ordinary Higgs. 

We consider that the existence of $\nu_R$ is quite plausible because all GUT groups larger than SU(5) require them. In
particular the fact that $\nu_R$ completes the representation 16 of SO(10): 16=$\bar 5$+10+1, so that all fermions of each
family are contained in a single representation of the unifying group, is too impressive not to be significant. At least as a
classification group SO(10) must be of some relevance. Thus in the following we assume that there are both $\nu_R$ and lepton
number violation. With these assumptions the see-saw mechanism \cite{ss} is possible. In its simplest form it arises as
follows. Consider the mass terms in the lagrangian corresponding to Dirac and RR Majorana masses (for the time being we
consider LL Majorana mass terms as comparatively negligible):
\beq {\cal L}=-{\ov R} m_D L'+\frac{1}{2}\ov{R} M \ov{R}^T+~ h.c.\label{lag}
\eeq  For notational simplicity we denoted $\nu_L$ and $\nu_R$ by $L'$ and $R$, respectively (the prime defining the $\nu$
flavour basis, see eq. (\ref{U})).  The 3 by 3 matrices $m_D$ and $M$ are the  Dirac and Majorana mass matrices in flavour
space  ($M$ is symmetric, $M=M^T$, while $m_D$ is, in general, non hermitian and non symmetric). We expect the eigenvalues of
$M$ to be of order $M_{GUT}$ or more because RR Majorana masses are $SU(3)\times SU(2)\times U(1)$ invariant,  hence
unprotected and naturally of the order of the cutoff of the low-energy theory.  Since all $\nu_R$ are very heavy we can
integrate them away.  For this purpose we write down the equations of motion for $\bar R$ in the static limit, $i.e.$
neglecting their kinetic terms:
\beq -\frac{\partial {\cal L}}{\partial \bar R}=m_D L'- M\bar R^T= 0\label{eulag}
\eeq From this, by solving for $\bar R^T$ and transposing,  we obtain:
\beq
\bar R=L'^T m_D^T M^{-1}\label{R}
\eeq We now replace in the lagrangian, eq. (\ref{lag}), this expression for $\bar R$ and we get:
\beq {\cal L}=-{1\over 2} L'^T m_{\nu} L'~~~~~,
\eeq where the resulting neutrino mass matrix is:
\beq m_{\nu}=m_D^T M^{-1}m_D
\eeq This is the well known see-saw mechanism \cite{ss} result: the light neutrino masses are quadratic in the Dirac masses and
inversely proportional to the large Majorana mass.  If some $\nu_R$ are massless or light they would not be integrated away but
simply added to the light neutrinos. Notice that the above results hold true for any number
$n$ of heavy neutral fermions 
$R$ coupled to the 3 known neutrinos. In this more general case $M$ is an $n$ by $n$ symmetric matrix and the coupling between
heavy and light fields is described by the rectangular $n$ by 3 matrix $m_D$. 

Here we assumed that the additional non renormalizable terms from $O_5$ are comparatively negligible, otherwise they should
simply be added. After elimination of the heavy right-handed fields, at the level of the effective low energy theory, the two
types of terms are equivalent. In particular they have identical transformation properties under a chiral change of basis in
flavour space. The difference is, however, that in the see-saw mechanism, the Dirac matrix $m_D$ is presumably related to
ordinary fermion masses because they are both generated by the Higgs mechanism and both must obey GUT-induced constraints. Thus
if we assume the see-saw mechanism more constraints are implied. In particular we are led to the natural hypothesis that
$m_D$ has a largely dominant third family eigenvalue in analogy to $m_t$, $m_b$ and $m_{\tau}$ which are by far the largest
masses among $u$ quarks, $d$ quarks and charged leptons. Once we accept that $m_D$ is hierarchical it is very difficult to
imagine that the effective light neutrino matrix, generated by the see-saw mechanism, could have eigenvalues very close in
absolute value.

\section{The Neutrino Mixing Matrix}

Given the definition of the mixing matrix $U$ in eq. (\ref{U}) and the transformation properties of the effective light
neutrino mass matrix $m_{\nu}$:
\bea  L'^Tm_{\nu}L'&= &L^TU^Tm_{\nu}UL\nonumber\\  U^Tm_{\nu}U& = &{\rm Diag}[e^{i\phi_1}m_1,e^{i\phi_2}m_2,m_3]\equiv m_{diag}
\label{tr}
\eea  we obtain the general form of $m_{\nu}$:
\beq m_{\nu}=U m_{diag} U^T\label{gen}
\eeq The matrix $U$ can be parameterized in terms of three mixing angles and one phase, exactly as for the quark mixing matrix
$V_{CKM}$. In addition we have the two phases $\phi_1$ and $\phi_2$ that are present because of the Majorana nature of
neutrinos. Thus, in general, 9 parameters are added to the SM when non vanishing neutrino masses are included: 3 eigenvalues, 3
mixing angles and 3 CP violating phases \cite{cp1,cp2}.

Maximal atmospheric neutrino mixing and the requirement that the electron neutrino does not participate in the atmospheric
oscillations, as indicated by the Superkamiokande
\cite{SK} and Chooz
\cite{Chooz} data, lead directly to the following structure of the
$U_{fi}$ ($f=e,\mu,\tau$, $i=1,2,3$) mixing matrix, apart from sign convention redefinitions 
%(here we are not interested in CP violation effects: all matrices are taken real)
\beq  U_{fi}= 
\left[\matrix{ c&-s&0 \cr s/\sqrt{2}&c/\sqrt{2}&-1/\sqrt{2}\cr s/\sqrt{2}&c/\sqrt{2}&+1/\sqrt{2}     } 
\right] ~~~~~.
\label{ufi}
\eeq  This result is obtained by a simple generalization of the analysis of ref. \cite{bar,Bal} to the case of arbitrary solar
mixing angle ($s\equiv\sin{\theta_{sun}}$, $c\equiv\cos{\theta_{sun}}$):
$c=s=1/\sqrt{2}$ for maximal solar mixing (e.g. for vacuum oscillations $\sin^2{2\theta_{sun}}\sim 0.75$) , while
$\sin^2{2\theta_{sun}}\sim 4s^2\sim 5.5\cdot 10^{-3}$ for the small angle MSW \cite{MSW} solution. The vanishing of
$U_{e3}$ guarantees that 
$\nu_e$ does not participate in the atmospheric oscillations and the relation
$|U_{\mu3}|=|U_{\tau3}|=1/\sqrt{2}$ implies maximal mixing for atmospheric neutrinos.  Also, in the limit $U_{e3}=0$, all CP
violating effects vanish and we can neglect the additional phase parameter generally present in $U_{fi}$: the matrix $U$ is
real and orthogonal and equal to the product of a rotation by $\pi/4$ in the 23 plane times a rotation in the 12 plane:
\beq  U_{fi}= 
\left[\matrix{ 1&0&0 \cr 0&1/\sqrt{2}&-1/\sqrt{2}\cr 0&1/\sqrt{2}&1/\sqrt{2}} 
\right] ~
\left[\matrix{ c&-s&0 \cr s&c&0\cr 0&0&1} 
\right] ~~~~~.
\label{2rot}
\eeq  
Note that we are assuming only two frequencies, given by 
\beq\Delta_{sun}\propto m^2_2-m^2_1,~~~~~~~
\Delta_{atm}\propto m^2_3-m^2_{1,2}\label{fre}
\eeq The numbering 1,2,3 corresponds to our definition of the frequencies and in principle may not coincide with the family
index although this will be the case in the models that we favour. The effective light neutrino mass matrix is given by eq.
(\ref{gen}). We disregard the phases $\phi_{1,2}$ but in the following $m_{1,2}$ can be of either sign. 

For generic $s$, using eq. (\ref{gen}), one finds
\beq  m_{\nu}= \left[\matrix{ 2\epsilon&\delta&\delta\cr
\delta&\dd\frac{m_3}{2}+\epsilon_2&-\dd\frac{m_3}{2}+\epsilon_2\cr
\delta&-\dd\frac{m_3}{2}+\epsilon_2&\dd\frac{m_3}{2}+\epsilon_2     } 
\right]~~~~~, 
\label{mneu}    
\eeq  with 
\beq
\epsilon=(m_1 c^2+m_2 s^2)/2~~,~~~~~\delta=(m_1-m_2) c s/\sqrt{2}~~, ~~~~~\epsilon_2=(m_1 s^2+m_2 c^2)/2~~~~~.
\label{ede}    
\eeq We see that the existence of one maximal mixing and
$U_{e3}=0$ are the most important input that lead to the matrix form in eqs. (\ref{mneu}-\ref{ede}). The value of the solar
neutrino mixing angle can be left free. While the simple parametrization of the matrix $U$ in eq. (\ref{ufi}) is quite useful
to guide the search for a realistic pattern of neutrino mass matrices, it should not be taken too literally. In particular the
data do not exclude a non-vanishing $U_{e3}$ element. As already mentioned, the bound by Chooz
\cite{Chooz} amounts to  $|U_{e3}|\lappeq 0.2$. Thus neglecting
$|U_{e3}|$ with respect to
$s$ in eq. (\ref{ufi}) is not completely justified. Also note that in presence of a large hierarchy
$|m_3|\gg |m_{1,2}|$ the  effect of neglected parameters in  eq. (\ref{ufi}) can be enhanced by $m_3/m_{1,2}$ and produce
seizable corrections. A non vanishing
$U_{e3}$ term can lead to different $(m_\nu)_{12}$ and  $(m_\nu)_{13}$ terms. Similarly a deviation from maximal mixing
$U_{\mu 3}\not=U_{\tau 3}$ distorts the $\epsilon_2$ terms in the 23 sector of
$m_{\nu}$. Therefore, especially for a large hierarchy, there is more freedom in the small terms in order to construct a model
that fits the data than it is apparent from eq. (\ref{mneu}).

\section{Possible Mass Hierarchies and Approximate Textures}

Given the observed frequencies and our notation in eq. (\ref{fre}), there are three possible hierarchies of mass eigenvalues
\cite{us2}:
\bea
        {\rm A}& : & |m_3| >> |m_{2,1}| \nonumber\\
        {\rm B}& : & |m_1|\sim |m_2| >> |m_3| \nonumber\\
        {\rm C}& : & |m_1|\sim |m_2| \sim |m_3|
\label{abc}
\eea (in case A there is no prejudice on the $m_1$, $m_2$ relation). For B and C different subcases are then generated
according to the relative sign assignments for $m_{1,2,3}$. 

\begin{table}
\tcaption{Zeroth order form of the neutrino mass matrix  for double and single maximal mixing, 
according to the different possible hierarchies given in eq. (\ref{abc}).}
\small
\begin{center}
\begin{tabular}{|c|c|c|c|}   
\hline                         & & double & single \\ & $m_{diag}$ & maximal & maximal \\ & & mixing & mixing \\
\hline & & & \\ A & Diag[0,0,1] & 
$\left[
\matrix{ 0&0&0\cr 0&\um&-\um\cr 0&-\um&\um}
\right]$ &
$\left[
\matrix{ 0&0&0\cr 0&\um&-\um\cr 0&-\um&\um}
\right]$ \\ & & & \\
\hline & & & \\ B1 & Diag[1,-1,0] & 
$\left[
\matrix{ 0&\sq&\sq\cr
\sq&0&0\cr
\sq&0&0}
\right]$ &
$\left[
\matrix{ 1&0&0\cr 0&-\um&-\um\cr 0&-\um&-\um}
\right]$ \\ & & & \\
\hline & & & \\ B2 & Diag[1,1,0] &
$\left[
\matrix{ 1&0&0\cr 0&\um&\um\cr 0&\um&\um}
\right]$ &
$\left[
\matrix{ 1&0&0\cr 0&\um&\um\cr 0&\um&\um}
\right]$ \\ & & & \\
\hline & & & \\ C0 & Diag[1,1,1] & 
$\left[
\matrix{ 1&0&0\cr 0&1&0\cr 0&0&1}
\right]$ &
$\left[
\matrix{ 1&0&0\cr 0&1&0\cr 0&0&1}
\right]$ \\ & & & \\
\hline & & & \\ C1 & Diag[-1,1,1] & 
$\left[
\matrix{ 0&-\sq&-\sq\cr -\sq&\um&-\um\cr -\sq&-\um&\um}
\right]$ &
$\left[
\matrix{ -1&0&0\cr 0&1&0\cr 0&0&1}
\right]$ \\ & & & \\
\hline & & & \\ C2 & Diag[1,-1,1] & 
$\left[
\matrix{ 0&\sq&\sq\cr
\sq&\um&-\um\cr
\sq&-\um&\um}
\right]$ &
$\left[
\matrix{ 1&0&0\cr 0&0&-1\cr 0&-1&0}
\right]$ \\ & & & \\
\hline & & & \\ C3 & Diag[1,1,-1] & 
$\left[
\matrix{ 1&0&0\cr 0&0&1\cr 0&1&0}
\right]$ &
$\left[
\matrix{ 1&0&0\cr 0&0&1\cr 0&1&0}
\right]$ \\ & & & \\
\hline                        
\end{tabular} 
\end{center}
\end{table}
For each case we can set to zero the small masses and mixing angles and find the effective light neutrino matrices which are
obtained both for double and single maximal mixing. Note that here we are working in the basis where the charged lepton masses
are diagonal, and approximately given by
$m_l={\rm Diag}[0,0,m_{\tau}]$. For model building one has to arrange both the charged lepton and the neutrino mass matrices so
that the neutrino results coincide with those given here after diagonalization of charged leptons.  For example, in case A,
$m_{diag}={\rm Diag}[0,0,m_3]$ and we obtain
\beq A:~~~m_{diag}={\rm Diag}[0,0,1]~m_3~~~~~~\rightarrow~~~~~~ m_{\nu}/m_3= \left[\matrix{ 0&0&0\cr
0&\frac{1}{2}&-\frac{1}{2}\cr 0&-\frac{1}{2}&\frac{1}{2}    } \right]~~~~~. 
\label{a}
\eeq In this particular case the results are the same for double and single maximal mixing. Note that the signs correspond to
the phase convention adopted in eq. (\ref{ufi}). If one prefers all signs to be positive it is sufficient to invert the sign of
the third row of the matrix $U$ in eq. (\ref{ufi}). We can similarly proceed in the other cases and we obtain the results in
table I (where the overall mass scale was dropped).

We recall that, once a solution to the solar neutrino problem is chosen and a set of suitable small perturbation terms is
introduced, oscillation phenomena are unable to distinguish between the cases $A$, $B$ and $C$ displayed in table I. However,
from the model building point of view, each texture in table I represents an independent possibility in a zeroth order
approximation. Moreover, table I could be generalized by allowing general phase differences among the 3 neutrino  masses as
shown in eq. (\ref{tr}). For instance, for double maximal mixing and $|m_1|=|m_2|=|m_3|$ we find the general texture
\beq  {m_{\nu}\over m_3}= {1\over 2}
\left[\matrix{ p_{12} & \dd{m_{12}\over\sqrt{2}} & \dd{m_{12}\over\sqrt{2}} \cr
\dd{m_{12}\over\sqrt{2}} & 1+\dd{p_{12}\over 2}     & -1+\dd{p_{12}\over 2}\cr
\dd{m_{12}\over\sqrt{2}} & -1+\dd{p_{12}\over 2}   &  1+\dd{p_{12}\over 2}} 
\right]~~~~~, 
\label{genf}    
\eeq  where $p_{12}(m_{12})\equiv e^{i\phi_1}+(-)e^{i\phi_2}$.  When $p_{12}=0$, eq. (\ref{genf}) reproduces the cases C1 and
C2 of table I, while for $m_{12}=0$ C0 and C3 are obtained.

\section{The Case of Nearly Degenerate Neutrino Masses}

The configurations B) and C) imply a very precise near degeneracy of squared masses. For example, the case C) is the only one
that could in principle accommodate neutrinos as hot dark matter together with solar and atmospheric neutrino oscillations. We
think that it is not at all clear at the moment that a hot dark matter component is really needed
\cite{kra} but this could be a reason in favour of the fully degenerate solution. Then the common mass should be around 1-3 eV.
The solar frequency could be given by a small 1-2 splitting, while the atmospheric frequency could be given by a still small
but much larger 1,2-3 splitting.  A strong constraint arises in this case from the non observation of neutrino-less double beta
decay which requires that the $ee$ entry of $m_{\nu}$ must obey
$|(m_{\nu})_{ee}|\leq 0.2~{\rm eV}$ \cite{dbeta}. Perhaps this bound can somewhat be relaxed if the error from the
theoretical ambiguities on nuclear matrix elements is increased. As observed in ref.
\cite{GG}, if this bound is below $1~\eV$, it can only be  satisfied if bimixing is realized (that is double maximal mixing,
with solar neutrinos explained by the VO solution or may be by the large angle MSW solution). The reason is that, as seen from eqs.
(\ref{mneu},\ref{ede}),
$(m_{\nu})_{ee}=m_1c^2+m_2s^2$. Now
$s^2$ is either very small or very close to $c^2$. For the required cancellation one needs opposite signs for $m_1$ and
$m_2$ and comparable values of
$c^2$ and $s^2$, hence nearly maximal solar mixing. 

A problem with this type of solution is that we would need a relative splitting
$|\Delta m/m|\sim
\Delta m^2_{atm}/2m^2\sim 10^{-3}-10^{-4}$ and a much smaller one for solar neutrinos especially if explained by vacuum
oscillations:
$|\Delta m/m|\sim 10^{-10}-10^{-11}$. As mentioned above we consider it unplausible that starting from hierarchical Dirac
matrices we end up via the see-saw mechanism into a nearly perfect degeneracy of squared masses. Thus models with degenerate
neutrinos  could only be natural if the dominant contributions directly arise from non renormalizable operators like $O_5$ in
eq. (\ref{O5}) because they are apriori unrelated to other fermion mass terms.  The degeneracy of neutrinos should be
guaranteed by some slightly broken symmetry.  Notice however that, even if arranged at the GUT scale, it is
doubtful that such a precise degeneracy could be stable against renormalization group corrections when running down at low
energy unless it is protected by a suitable symmetry \cite{ello}.  Models based on discrete or continuous symmetries have been proposed. For example in the models of ref.
\cite{BHKR} the symmetry is SO(3). In the unbroken limit neutrinos are degenerate and charged leptons are massless. When the
symmetry is broken the charged lepton masses are much larger than neutrino splittings because the former are first order while
the latter are second order in the electroweak symmetry breaking.

A model which is simple to describe but difficult to derive in a natural way is one \cite{fri,wet} where up quarks, down quarks
and charged leptons have "democratic" mass matrices, with all entries equal (in first approximation):
\beq m_D^u,~m_D^d,~m_D^l \propto \frac{1}{3} \left[
\matrix{ 1&1&1\cr 1&1&1\cr 1&1&1}
\right] \rightarrow \left[
\matrix{ 0&0&0\cr 0&0&0\cr 0&0&1}
\right]\label{dem}
\eeq where we have also indicated the diagonal form with two vanishing eigenvalues. In this limit the CKM matrix is the
identity, because $u$ and $d$ quarks are diagonalized by the same matrix (recall that $V_{CKM}=U^\dagger_u U_d$).  This matrix
$U$ can be chosen with $U_{e3}=0$ and is given by (note the analogy with the quark model eigenvalues $\pi^0$,
$\eta$ and
$\eta'$):
\beq U= \left[
\matrix{ 1/\sqrt{2}&-1/\sqrt{2}&0\cr 1/\sqrt{6}&1/\sqrt{6}&-2/\sqrt{6}\cr 1/\sqrt{3}&1/\sqrt{3}&1/\sqrt{3}}
\right]~~~~~.
\label{ufri}
\eeq

The mass matrix of eq. (\ref{dem}) is invariant under a discrete
$S_{3L}\times S_{3R}$ permutation symmetry. The same requirement leads to the general neutrino mass matrix:
\beq m_\nu \propto a \left[
\matrix{ 1&0&0\cr 0&1&0\cr 0&0&1}
\right] +b
\left[
\matrix{ 1&1&1\cr 1&1&1\cr 1&1&1}
\right]
\label{dem1}~~~~~,
\eeq the two independent invariants being allowed by the Majorana nature of the light neutrinos. If $b$ vanishes the neutrinos
are degenerate. In the presence of small terms that break the permutation symmetry this degeneracy is removed and the neutrino
mixing matrix may remain very close to $U$ in eq. (\ref{ufri}). At the same time, the lightest quarks and charged leptons may
acquire a non-vanishing mass. Notice that the atmospheric neutrino mixing is nearly maximal:
$\sin^2{2\theta}=8/9$, while the solar mixing angle is maximal.

An intermediate possibility between the case of degenerate neutrino masses and the case of hierarchical ones is the situation
$B$ in eq. (\ref{abc}). With two almost degenerate heavier states and a nearly massless neutrino we cannot reach a mass range
interesting for cosmological purposes. However, since the experimentally accessible quantities are the squared mass
differences, case $B$ remains an open possibility. The small terms required to go beyond the zeroth order approximation can be
controlled by a spontaneously broken flavour symmetry. For instance, working only with the light neutrinos, the $U(1)_Q$ charge
$Q\equiv(L_e-L_\mu-L_\tau)$ allows for a texture of the kind  B1, double maximal mixing, at leading order \cite{hall}. The
non-vanishing  entries $(m_\nu)_{12}$ and $(m_\nu)_{13}$ are  expected to be of the same order, although not necessarily equal.
The vanishing entries can be filled by the ratio of the VEV
$\langle\varphi\rangle$ of a scalar field carrying two units of $Q$ and a mass scale $M$ providing the cut-off to the
low-energy theory:
\beq  m_\nu=m
\left[\matrix{
\epsilon&1&1\cr 1&\epsilon&\epsilon\cr 1&\epsilon&\epsilon} 
\right]~~~~~, 
\label{B1}
\eeq where $\epsilon=\langle\varphi\rangle/M$ and only the order-of-magnitudes are indicated. For values of $\epsilon$ smaller
than 1 one obtains a small perturbation of the zeroth order texture. The mixing matrix has a large, not necessarily maximal,
mixing angle in the 23 sector, a nearly maximal mixing angle in the 12 sector and a mixing angle of order $\epsilon$ in the 13
sector \cite{ghos}. The mass parameter $m$ should be  close to $10^{-1}-10^{-2}$ eV to provide the frequency required by the atmospheric
oscillations. Finally,
$m_2^2-m_1^2\sim m^2\epsilon$. In this model the VO solution to the solar neutrino deficit would require a very tiny
breaking term, $\epsilon\sim {\rm O}(10^{-7})$.
  
\section{The Case of Hierarchical Neutrino Masses}

We now discuss models of type A with large effective light neutrino mass splittings and large mixings. In general large
splittings correspond to small mixings because normally only close-by states are strongly mixed. In a 2 by 2 matrix context the
requirement of large splitting and large mixings leads to a condition of vanishing determinant. For example the matrix
\beq m\propto 
\left[\matrix{ x^2&x\cr x&1    } 
\right]~~~~~. 
\label{md0}
\eeq has eigenvalues 0 and $1+x^2$ and for $x$ of O(1) the mixing is large. Thus, in the limit of neglecting small mass terms of
order $m_{1,2}$, the demands of large atmospheric neutrino mixing and dominance of $m_3$ translate into the condition that the 2
by 2 subdeterminant 23 of the 3 by 3 mixing matrix vanishes. The problem is to show that this approximate vanishing can be
arranged in a natural way without fine tuning. We have discussed suitable possible mechanisms in our papers
\cite{us1,us2,us3}. We in particular favour a class of models where, in the limit of neglecting terms of order
$m_{1,2}$ and, in the basis where charged leptons \underline{are diagonal}, the Dirac matrix $m_D$, defined by $\bar R m_D L$,
takes the approximate form:
\beq m_D\propto 
\left[\matrix{ 0&0&0\cr 0&0&0\cr 0&x&1    } 
\right]~~~~~. 
\label{md00}
\eeq This matrix has the property that for a generic Majorana matrix $M$ one finds:
\beq m_{\nu}=m^T_D M^{-1}m_D\propto 
\left[\matrix{ 0&0&0\cr 0&x^2&x\cr 0&x&1    } 
\right]~~~~~. 
\label{mn0}
\eeq The only condition on $M^{-1}$ is that the 33 entry is non zero. It is important for the following discussion to observe
that $m_D$ given by eq. (\ref{md00}) under a change of basis transforms as $m_D\to V^{\dagger} m_D U$ where $V$ and $U$ rotate
the right and left fields respectively. It is easy to check that in order to make $m_D$ diagonal we need large left mixings.
More precisely $m_D$ is diagonalized by taking $V=1$ and $U$ given by
\beq U=  \left[
\matrix{ c&-s&0\cr sc_{\gamma}&cc_{\gamma}&-s_{\gamma}\cr ss_{\gamma}&cs_{\gamma}&c_{\gamma}  } 
\right]~~~~~, 
\label{uu}
\eeq with 
\beq s_{\gamma}=-x/r~~,~~~~~c_{\gamma}=1/r~~,~~~~~r=\sqrt{1+x^2}~~~~~.
\label{scr}
\eeq The matrix $U$ is directly the neutrino mixing matrix. The mixing angle for atmospheric neutrino oscillations is given by:
\beq
\sin^2{2\theta}=4s^2_{\gamma}c^2_{\gamma}=\frac{4 x^2}{(1+x^2)^2}~~~~~.
\label{sin}
\eeq Thus the bound $\sin^2{2\theta}\gappeq0.9$ translates into  $0.7\lappeq |x|\lappeq 1.4$. It is interesting to recall that
in refs. \cite{ellis,hagi} it was shown that the mixing angle can be amplified by the running from a large mass scale down to
low energy.

We have seen that, in order to explain in a natural way widely split light neutrino masses together with large mixings, we need
an automatic vanishing of the 23 subdeterminant. This in turn is most simply realized by allowing some large left-handed mixing
terms in the Dirac neutrino matrix. By left-handed mixing we mean non diagonal matrix elements that can only be eliminated by a
large rotation of the left-handed fields. Thus the question is how to reconcile large left-handed mixings in the leptonic
sector with the observed near diagonal form of $V_{CKM}$, the quark mixing matrix. Strictly speaking, since
$V_{CKM}=U^{\dagger}_u U_d$, the individual matrices $U_u$ and $U_d$ need not be near diagonal, but
$V_{CKM}$ does, while the analogue for leptons apparently cannot be near diagonal. However nothing forbids for quarks that, in
the basis where $m_u$ is diagonal, the $d$ quark matrix has large non diagonal terms that can be rotated away by a pure
right-handed rotation. We suggest that this is so and that in some way right-handed mixings for quarks correspond to
left-handed mixings for leptons.

In the context of (Susy) SU(5) \cite{ross} there is a very attractive hint of how the present mechanism can be realized. In the
$\bar 5$ of SU(5) the $d^c$ singlet appears together with the lepton doublet $(\nu,e)$. The $(u,d)$ doublet and $e^c$ belong to
the 10 and $\nu^c$ to the 1 and similarly for the other families. As a consequence, in the simplest model with mass terms
arising from only Higgs pentaplets, the Dirac matrix of down quarks is the transpose of the charged lepton matrix:
$m^d_D=(m^l_D)^T$. Thus, indeed, a large mixing for right-handed down quarks corresponds to a large left-handed mixing for
charged leptons.  At leading order we may have:
\beq m^d_D=(m^l_D)^T=
\left[
\matrix{ 0&0&0\cr 0&0&1\cr 0&0&1}
\right]v_d
\eeq In the same simplest approximation with  5 or $\bar 5$ Higgs, the up quark mass matrix is symmetric, so that left and
right mixing matrices are equal in this case. Then small mixings for up quarks and small left-handed mixings for down quarks
are sufficient to guarantee small $V_{CKM}$ mixing angles even for large $d$ quark right-handed mixings.  If these small
mixings are neglected, we expect:
\beq m^u_D=
\left[
\matrix{ 0&0&0\cr 0&0&0\cr 0&0&1}
\right]v_u~~~.
\eeq When the charged lepton matrix is diagonalized the large left-handed mixing of the charged leptons is transferred to the
neutrinos. Note that in SU(5) we can diagonalize the $u$ mass matrix by a rotation of the fields in the 10, the Majorana matrix
$M$ by a rotation of the 1 and the effective light neutrino matrix
$m_\nu$ by a rotation of the $\bar 5$. In this basis the $d$ quark mass matrix fixes $V_{CKM}$ and the charged lepton mass
matrix fixes neutrino mixings. It is well known that a model where the down and the charged lepton matrices are exactly the
transpose of one another cannot be exactly true because of the $e/d$ and
$\mu/s$ mass ratios. It is also known that one remedy to this problem is to add some Higgs component in the 45 representation
of SU(5) \cite{jg}. A different solution \cite{eg} will be described later. But the symmetry under transposition can still be a
good guideline if we are only interested in the order of magnitude of the matrix entries and not in their exact values.
Similarly, the Dirac neutrino mass matrix
$m_D$ is the same as the up quark mass matrix in the very crude model where the Higgs pentaplets come from a pure 10
representation of SO(10):
$m_D=m^u_D$. For $m_D$ the dominance of the third family eigenvalue  as well as a near diagonal form could be an order of
magnitude remnant of this broken symmetry. Thus, neglecting small terms, the neutrino Dirac matrix in the basis where charged
leptons are diagonal could be directly obtained in the form of eq. (\ref{md00}).

\section{An Example with Horizontal Abelian Charges}
 
We give here an explicit example \cite{us3} of the mechanism under discussion in the framework of a unified Susy $SU(5)$ theory
with an additional 
$U(1)_F$ flavour symmetry \cite{fro}. This model is to be taken as merely indicative, in that some important problems, like,
for example, the cancellation of chiral anomalies are not tackled here. But we find it impressive that the general pattern of
all what we know on fermion masses and mixings is correctly reproduced at the level of orders of magnitude.  We regard the
present model as a low-energy effective theory valid at energies close to $M_{GUT}\ll M_{Pl}$. We can think to obtain  it by
integrating out the heavy modes from an unknown underlying fundamental theory defined at an energy scale  close to $M_{Pl}$. From 
this point of view the gauge anomalies generated by the light supermultiplets listed below  can be compensated by another
set of supermultiplets with masses above $M_{GUT}$, already eliminated from the  low-energy theory. In particular, we assume
that these  additional supermultiplets are vector-like with respect to
$SU(5)$ and chiral with respect to $U(1)_F$. Their masses are then naturally expected to be of the order of the $U(1)_F$
breaking scale, which, in the following discussion, turns out to be near $M_{Pl}$. It is possible to check explicitly the 
possibility of canceling the gauge anomalies in this way but, due to our ignorance about the fundamental theory,  it is not
particularly instructive to illustrate the details here. In this model the known generations of quarks and leptons are
contained in  triplets
$\Psi_{10}^a$ and
$\Psi_{\bar 5}^a$, $(a=1,2,3)$ transforming as $10$ and ${\bar 5}$ of $SU(5)$, respectively. Three more $SU(5)$ singlets
$\Psi_1^a$ describe the right-handed neutrinos. We assign to these fields the following $F$-charges:
\bea
\Psi_{10}     & \sim & (3,2,0) \label{c10}\\
\Psi_{\bar 5} & \sim & (3,0,0) \label{c5b}\\
\Psi_1        & \sim & (1,-1,0) \label{c1}
\eea   We start by discussing the Yukawa coupling allowed by $U(1)_F$-neutral  Higgs multiplets $\varphi_5$ and
$\varphi_{\bar5}$ in the $5$ and ${\bar 5}$ $SU(5)$ representations and by a pair $\theta$ and
${\bar\theta}$ of $SU(5)$ singlets with $F=1$ and $F=-1$, respectively. 

In the quark sector we obtain 
\footnote{In eq. (\ref{mquark}) the entries denoted by 1 in $m_D^u$ and $m_D^d$  are not necessarily equal. As usual, such a
notation allows  for O(1) deviations.}
:
\beq m_D^u=(m_D^u)^T=
\left[
\matrix{
\lambda^6&\lambda^5&\lambda^3\cr
\lambda^5&\lambda^4&\lambda^2\cr
\lambda^3&\lambda^2&1}
\right]v_u~~,~~~~~~~ m_D^d=
\left[
\matrix{
\lambda^6&\lambda^5&\lambda^3\cr
\lambda^3&\lambda^2&1\cr
\lambda^3&\lambda^2&1}
\right]v_d~~,
\label{mquark}
\eeq  from which we get the order-of-magnitude relations:
\bea m_u:m_c:m_t & = &\lambda^6:\lambda^4:1 \nonumber\\ m_d:m_s:m_b & = &\lambda^6:\lambda^2:1
\eea and 
\beq V_{us}\sim \lambda~,~~~~~ V_{ub}\sim \lambda^3~,~~~~~ V_{cb}\sim \lambda^2~.
\eeq 

Here $v_u\equiv\langle \varphi_5 \rangle$, 
$v_d\equiv\langle \varphi_{\bar 5} \rangle$  and $\lambda$ denotes the ratio between the vacuum expectation value of
${\bar\theta}$ and an ultraviolet cut-off identified with the Planck mass $M_{Pl}$: $\lambda\equiv\langle
{\bar\theta}\rangle/M_{Pl}$. To correctly reproduce the observed quark mixing angles, we  take $\lambda$ of the order of the
Cabibbo angle. For non-negative $F$-charges, the elements of  the quark mixing matrix $V_{CKM}$ depend only on the charge
differences  of the left-handed quark doublet \cite{fro}. Up to a constant shift, this defines the choice in eq. (\ref{c10}).
Equal $F$-charges for $\Psi_{\bar 5}^{2,3}$ (see eq. (\ref{c5b})) are then required to fit
$m_b$ and $m_s$. We will comment on the lightest quark masses later on.

At this level, the mass matrix for the charged leptons  is the transpose of $m_D^d$:
\beq m_D^l=(m_D^d)^T
%=\left[
%\matrix{
%\lambda^6&\lambda^3&\lambda^3\cr
%\lambda^5&\lambda^2&\lambda^2\cr
%\lambda^3&1&1
%\right]
\eeq and we find:
\beq m_e:m_\mu:m_\tau  = \lambda^6:\lambda^2:1 
\eeq The O(1) off-diagonal entry of $m_D^l$ gives rise to a large left-handed  mixing in the 23 block which corresponds to a
large right-handed mixing in the $d$ mass matrix. In the neutrino sector, the Dirac and Majorana mass matrices are given by:
\beq m_D=
\left[
\matrix{
\lambda^4&\lambda&\lambda\cr
\lambda^2&\lambda'&\lambda'\cr
\lambda^3&1&1}
\right]v_u~~,~~~~~~~~ M=
\left[
\matrix{
\lambda^2&1&\lambda\cr 1&\lambda'^2&\lambda'\cr
\lambda&\lambda'&1}
\right]{\bar M}~~,
\eeq where $\lambda'\equiv\langle\theta\rangle/M_{Pl}$ and ${\bar M}$  denotes the large mass scale associated to the
right-handed neutrinos: ${\bar M}\gg v_{u,d}$.

After diagonalization of the charged lepton sector and after integrating out the heavy right-handed neutrinos we obtain the
following neutrino mass matrix in the low-energy effective theory:
\beq m_\nu=
\left[
\matrix{
\lambda^6&\lambda^3&\lambda^3\cr
\lambda^3&1&1\cr
\lambda^3&1&1}\right]{v_u^2\over {\bar M}}
\label{mnu}
\eeq where we have taken $\lambda\sim\lambda'$. The O(1) elements in the 23 block are produced by combining the large 
left-handed mixing induced by the charged lepton sector and the large left-handed mixing in $m_D$. A crucial property of
$m_\nu$ is that, as a result of the sea-saw mechanism and of the specific $U(1)_F$ charge assignment, the determinant of the 23
block is $\underline{automatically}$ of $O(\lambda^2)$ (for this the presence of negative charge values, leading to the
presence of both $\lambda$ and $\lambda'$ is essential \cite{us2}).

It is easy to verify that the eigenvalues of $m_\nu$ satisfy the relations:
\beq m_1:m_2:m_3  = \lambda^4:\lambda^2:1~~.
\eeq The atmospheric neutrino oscillations require 
$m_3^2\sim 10^{-3}~{\rm eV}^2$. From eq. (\ref{mnu}), taking $v_u\sim 250~{\rm GeV}$, the mass scale ${\bar M}$ of the heavy
Majorana neutrinos turns out to be close to the unification scale, 
${\bar M}\sim 10^{15}~{\rm GeV}$. The squared mass difference between the lightest states is  of $O(\lambda^4)~m_3^2$,
appropriate to the MSW solution  to the solar neutrino problem. Finally, beyond the large mixing in the 23 sector,
corresponding to $s_\gamma\sim c_\gamma$ in eq. (\ref{uu}),
$m_\nu$  provides a mixing angle $s \sim (\lambda/2)$ in the 12 sector, close to the range preferred by the small angle MSW
solution. In general $U_{e3}$ is non-vanishing, of $O(\lambda^3)$.

In general, the charge assignment under 
$U(1)_F$ allows for non-canonical kinetic terms that represent an additional source of mixing. Such terms are allowed by the
underlying flavour symmetry and it would be unnatural to tune them to the canonical form. The results quoted up to now  remain
unchanged after including the effects related to the most general kinetic terms, via appropriate  rotations and rescaling in
the flavour space (see also ref. \cite{lns}). 

Obviously, the order of magnitude description offered by this model is not intended to account for all the details of fermion
masses. Even neglecting the parameters associated with the $CP$ violating observables, some of the relevant observables are
somewhat marginally reproduced.   For instance we obtain $m_u/m_t\sim \lambda^6$  which is perhaps too large. However we find
it remarkable that in such a simple scheme most of the 12 independent  fermion masses and the 6 mixing angles turn out to have
the correct order of magnitude. Notice also that our model prefers large values of
$\tan\beta\equiv v_u/v_d$. This is a consequence of the equality $F(\Psi_{10}^3)=F(\Psi_{\bar 5}^3)$ (see eqs. (\ref{c10}) and
(\ref{c5b})). In this case the Yukawa couplings of top and bottom quarks  are expected to be of the same order of magnitude,
while the large
$m_t/m_b$ ratio is attributed to $v_u \gg  v_d$ (there may be factors O(1) modifying these considerations, of course). We
recall here that in supersymmetric grand unified models large values of $\tan\beta$  are one possible solution to the problem
of reconciling the boundary  condition $m_b=m_\tau$ at the  GUT scale with the low-energy data \cite{largetb}.  Alternatively,
to keep $\tan\beta$ small, one could  suppress $m_b/m_t$ by adopting different $F$-charges for the 
$\Psi_{\bar 5}^3$ and $\Psi_{10}^3$.

Additional contributions to flavour changing processes (both in the quark and in the lepton \cite{vis2} sectors) and to
$CP$ violating observables are generally expected in a supersymmetric grand unified theory. However, a reliable estimate of the
corresponding effects would require a much more detailed definition of the theory than attempted here. Crucial ingredients such
as the mechanism of supersymmetry breaking and its transmission to the observable sector have been ignored in the present note.
We are implicitly assuming that the omission of this aspect of the flavour problem does not substantially alter our discussion. 

A common problem of all $SU(5)$ unified theories based on a minimal higgs structure is represented by the relation
$m_D^l=(m_D^d)^T$ that, while leading to the successful $m_b=m_\tau$ boundary condition at the GUT scale, provides the wrong
prediction
$m_d/m_s=m_e/m_\mu$ (which, however, is an acceptable order of magnitude equality). We can easily overcome this problem and
improve the picture \cite{eg} by introducing an additional supermultiplet
${\bar\theta}_{24}$ transforming in the adjoint representation of $SU(5)$ and  possessing a negative $U(1)_F$ charge,
$-n~~(n>0)$.  Under these conditions, a positive
$F$-charge $f$ carried by the matrix elements 
$\Psi_{10}^a \Psi_{\bar 5}^b$ can be compensated  in several different ways by monomials of the kind
$({\bar\theta})^p({\bar\theta}_{24})^q$, with
$p+n q=f$. Each of these possibilities represents an independent contribution to the down quark and charged lepton mass
matrices, occurring with an unknown coefficient of O(1). Moreover the product $({\bar\theta}_{24})^q \varphi_{\bar 5}$ contains
both the ${\bar 5}$ and the $\overline{45}$ $SU(5)$ representations,  allowing for a differentiation between the down quarks
and the charged leptons. The only, welcome, exceptions are  given by the O(1) entries that do not require any compensation and,
at the leading order, remain the same for charged leptons and  down quarks. This preserves the good
$m_b=m_\tau$ prediction. Since a perturbation of O(1) in the subleading matrix elements  is sufficient to cure the bad
$m_d/m_s=m_e/m_\mu$ relation, we can safely assume that 
$\langle{\bar\theta}_{24}\rangle/M_{Pl}\sim\lambda^n$, to preserve the correct order-of-magnitude predictions in the remaining
sectors. 

A general problem common to all models dealing with flavour is that of recovering the correct  vacuum structure by minimizing
the effective potential of the theory. It may be noticed that the presence of two multiplets $\theta$ and
${\bar \theta}$ with opposite $F$ charges could hardly be reconciled, without adding extra structure to the model, with a large
common VEV for these fields, due to possible analytic terms of the kind $(\theta {\bar \theta})^n$ in the superpotential
\cite{irges}.
%\footnote{We thank N. Irges for bringing our attention on this point.}.  
We find therefore instructive to explore the
consequences of allowing only the negatively charged ${\bar \theta}$ field in the theory.

It can be immediately recognized that, while the quark mass matrices of eqs. (\ref{mquark}) are unchanged, in the neutrino
sector the Dirac  and Majorana matrices get modified into:
\beq m_D=
\left[
\matrix{
\lambda^4&\lambda&\lambda\cr
\lambda^2&0&0\cr
\lambda^3&1&1}
\right]v_u~~,~~~~~~~~ M=
\left[
\matrix{
\lambda^2&1&\lambda\cr 1&0&0\cr
\lambda&0&1}
\right]{\bar M}~~.
\eeq The zeros are due to the analytic property of the superpotential that makes impossible to form the corresponding $F$
invariant by using ${\bar \theta}$ alone. These zeros should not be taken literally, as they will be eventually   filled by
small terms coming, for instance, from the diagonalization of the charged lepton mass matrix and from the transformation that
put the kinetic terms into canonical form. It is however interesting to work out, in first approximation, the case  of exactly
zero entries in $m_D$ and $M$, when forbidden by $F$.

The neutrino mass matrix obtained via see-saw from $m_D$ and $M$ has the same pattern as the one displayed in eq. (\ref{mnu}).
A closer inspection reveals that the determinant of the 23 block is identically zero, independently from
$\lambda$. This leads to the following pattern of masses:
\beq m_1:m_2:m_3  = \lambda^3:\lambda^3:1~~,~~~~~m_1^2-m_2^2 = {\rm O}(\lambda^9)~~.
\eeq Moreover the mixing in the 12 sector is almost maximal:
\beq {s\over c}={\pi\over 4}+{\rm O}(\lambda^3)~~.
\eeq For $\lambda\sim 0.2$, both the squared mass difference $(m_1^2-m_2^2)/m_3^2$  and $\sin^2 2\theta_{sun}$ are remarkably
close to the values  required by the vacuum oscillation solution to the solar neutrino problem. This property  remains
reasonably stable against the perturbations induced by small terms (of order $\lambda^5$) replacing the zeros, coming from the
diagonalization of the charged lepton sector  and by the transformations that render the kinetic terms canonical. We find quite
interesting that also the just-so solution, requiring  an intriguingly small mass difference and a bimaximal mixing, can be
reproduced, at least at the level of order of magnitudes, in the context of a "minimal" model of flavour compatible with
supersymmetric SU(5). In this case the role played by supersymmetry  is essential, a non-supersymmetric model with ${\bar
\theta}$ alone  not being distinguishable from the version with both
$\theta$ and ${\bar \theta}$, as far as low-energy flavour properties are concerned.

\section{Other Similar or Alternative Proposals}

Finally, let us compare our model with other recent proposals \cite{Large}. Textures for the effective neutrino mass matrix
similar to $m_\nu$ in eq. (\ref{mnu}) were derived in refs. \cite{ram,buch}, also in the context of an $SU(5)$ unified theory
with a $U(1)$ flavour symmetry. In these works, however, the O(1) entries of
$m_\nu$ are uncorrelated, due to the particular choice of U(1) charges. The diagonalization of such matrix, for generic O(1)
coefficients, leads to only one  light eigenvalue and to two heavy eigenvalues, of O(1), in units of 
$v_u^2/{\bar M}$. Then the required pattern $m_3\gg m_2\sim m_1$ has to be fixed by hand. On the contrary, in
 our model the desired pattern is automatic, since, as emphasized above, the determinant of the 23 block in $m_\nu$ is
vanishing at the leading  order. Other models in terms of $U(1)$ horizontal charges have been proposed in refs.
\cite{abelian,froggart,Bando,ellis,maro}. Clearly a large mixing for the light neutrinos can be provided in part by the
diagonalization of the charged lepton sector. As we have seen, in
$SU(5)$, the left-handed mixing  carried by charged leptons is expected to be, at least in first approximation, directly linked
to the right-handed mixing for the $d$ quarks and, as such, perfectly compatible with the available data. This possibility was
remarked, for instance, in refs. \cite{bere0,alb,bere,hagi} where the implementation was in terms of asymmetric textures, of
the Branco et al. type \cite{bra}, used as a general parametrization of the  existing data consistent with the constraints
imposed by the unification program. On the other hand, our model aims to a dynamical explanation of the flavour properties,
although in a simplified setting 
\footnote{Also ref. \cite{bere} suggests an horizontal $U(2)$ symmetry to justify the assumed textures.}~.

It is interesting to note that the mechanism discussed sofar can be embedded in an $SO(10)$ grand-unified theory in a rather
economic way \cite{alb}. The 33 entries of the fermion mass matrices can be obtained through the coupling ${\bf 16}_3 {\bf
16}_3 {\bf 10}_H$ among the fermions in the third generation, ${\bf 16}_3$, and a Higgs tenplet ${\bf 10}_H$. The two
independent VEVs of the tenplet $v_u$ and $v_d$ give mass, respectively, to $t/\nu_\tau$ and $b/\tau$. The keypoint to obtain
an asymmetric texture is the introduction of an operator of the kind ${\bf 16}_2 {\bf 16}_H {\bf 16}_3 {\bf 16}_H'$ . This
operator is thought to arise by integrating out an heavy {\bf 10} that couples both to ${\bf 16}_2 {\bf 16}_H$ and to ${\bf
16}_3 {\bf 16}_H'$. If the ${\bf 16}_H$ develops a VEV breaking $SO(10)$ down to $SU(5)$ at a large scale, then, in terms of
$SU(5)$ representations, we get an effective coupling of the kind ${\bf\ov{5}}_2 {\bf 10}_3 {\bf\ov{5}}_H$, with a coefficient
that can be of order one. This coupling contributes to the 23 entry of the down quark mass matrix  and to the 32 entry of the
charged lepton mass matrix, realizing the desired asymmetry.   To distinguish the lepton and quark sectors one can further
introduce  an operator of the form ${\bf 16}_i {\bf 16}_j {\bf 10}_H {\bf 45}_H$, $(i,j=2,3)$, with the VEV of the 
${\bf 45}_H$ pointing in the $B-L$ direction. Additional operators, still of the type 
${\bf 16}_i {\bf 16}_j {\bf 16}_H {\bf 16}_H'$ can contribute to the matrix elements of the first generation. The mass matrices
look like:
\beq m_D^u=
\left[
\matrix{ 0& 0& 0\cr 0& 0& \epsilon/3\cr 0&-\epsilon/3&1}
\right]v_u~~,~~~~~~~ m_D^d=
\left[
\matrix{ 0&\delta&\delta'\cr
\delta&0&\sigma+\epsilon/3\cr
\delta'&-\epsilon/3&1}
\right]v_d~~,
\label{mquark1}
\eeq 
\beq m_D=
\left[
\matrix{ 0& 0& 0\cr 0& 0& -\epsilon\cr 0& \epsilon&1}
\right]v_u~~,~~~~~~~ m_D^e=
\left[
\matrix{ 0&\delta&\delta'\cr
\delta&0&-\epsilon\cr
\delta'&\sigma+\epsilon&1}
\right]v_d~~.
\label{mquark2}
\eeq  They provide a good fit of the available data in the quarks and the charged lepton sector in terms of 5  parameters (one
of which is complex). In the neutrino sector one obtains a large $\theta_{23}$ mixing angle,
$\sin^2 2\theta_{12}\sim 6.6\cdot 10^{-3}$ eV$^2$ and $\theta_{13}$ of the same order of $\theta_{12}$. Mass squared
differences are sensitive to the details of the Majorana mass matrix.

Looking at models with three light neutrinos only, i.e. no sterile neutrinos, from a more general point of view, we stress that
in the above models the atmospheric neutrino mixing is considered large, in the sense of being of order one in some zeroth
order approximation. In other words it corresponds to off diagonal matrix elements of the same order of the diagonal ones,
although the mixing is not exactly maximal. The idea that all fermion mixings are small and induced by the observed smallness
of the non diagonal $V_{CKM}$  matrix elements is then abandoned. An alternative is to argue that perhaps what appears to
be large is not that large after all. The typical small parameter that appears in the mass matrices is $\lambda\sim
\sqrt{m_d/m_s}
\sim
\sqrt{m_{\mu}/m_{\tau}}\sim 0.20-0.25$. This small parameter is not so small that it cannot become large due to some peculiar
accidental enhancement: either a coefficient of about 3, or an exponent of the mass ratio which is less than $1/2$ (due for
example to a suitable charge assignment), or the addition in phase of an angle from the diagonalization of charged leptons and
an angle from neutrino mixing. One may like this strategy of producing a large mixing by stretching small ones if, for
example, he/she likes symmetric mass matrices, as from left-right symmetry at the GUT scale. In left-right symmetric models
smallness of left mixings implies that also right-handed mixings are small, so that all mixings tend to be small. Clearly this
set of models tend to favour moderate hierarchies and a single maximal mixing, so that the SA-MSW solution of solar neutrinos
is preferred. For example, consider the 23 submatrix only, for simplicity. Assume that at order zero the neutrino Dirac matrix
is given by 
\beq m_D\sim 
\left[
\matrix{ 0&0\cr 0&1}
\right]
\label{eq1}
\eeq For a generic Majorana matrix $M$ also the effective light neutrino matrix $m_{\nu}$ has the same structure in the 23
sector. If for charged leptons one has 
\beq m_l\sim 
\left[
\matrix{ \lambda^2&r \lambda\cr r \lambda&1}
\right]
\label{eq2}
\eeq with $r\sim {\rm O}(1)$ and $\lambda \sim {\rm O}(\sqrt{m_{\mu}/m_{\tau}})$ (in fact the eigenvalues are of order 1 and
$\lambda^2$, respectively), then the symmetric charged lepton matrix is diagonalized by a unitary matrix (the same for left and
right fields). In the basis where
$m_l$ is diagonal, we then have
\beq m_{\nu}\sim 
\left[
\matrix{ c&s\cr -s&c}
\right]
\left[
\matrix{ 0&0\cr 0&1}
\right]
\left[
\matrix{ c&-s\cr s&c}
\right]=
\left[
\matrix{ s^2& s c\cr s c&c^2}
\right]
\label{eq3}
\eeq As a result the 23 sub-determinant vanishes and $\sin{2\theta}\sim 2sc\sim 2r\lambda$ is large for $r\sim 2-3$. Models of
this type have been discussed in refs. \cite{Lola}.

To further discuss how a large neutrino mixing could be generated starting from symmetric matrices and small mixings for quarks
and leptons we can go, without loss of generality, to a basis where both the charged lepton Dirac mass matrix and the RR
Majorana matrix are diagonal. In fact, after diagonalization of the charged lepton Dirac mass matrix, we still have the freedom
of a change of basis for the right-handed neutrino fields, in that the right-handed charged lepton and neutrino fields, as
opposed to left-handed fields, are uncorrelated by the $SU(2)\bigotimes U(1)$ gauge symmetry. We can use this freedom to make
the Majorana matrix diagonal:
$M^{-1}=V^Td_MV$ with
$d_M={\rm Diag}[1/M_1,1/M_2,1/M_3]$. Then if we parametrize the matrix
$Vm_D$ by $z_{ab}$ we have:
\beq m_{\nu ab}=(m_D^TM^{-1}m_D)_{ab}=\sum_c \frac{z_{ca}z_{cb}}{M_c}.\label{cc}
\eeq From this expression we see that, while we can always arrange the twelve parameters $z_{ab}$ and $M_a$ to arbitrarily fix
the six independent matrix elements of $m_{\nu}$, case  A is special in that it can be approximately reproduced in two
particularly simple ways, without relying on precise cancellations among different terms: 
\begin{enumerate}
\item[i)] There are only two large entries in the $z$ matrix,
$|z_{c2}|\sim |z_{c3}|$, and the three eigenvalues $M_a$ are of comparable magnitude (or, at least, with a less pronounced
hierarchy than for the $z$ matrix elements). Then, the subdeterminant 23 vanishes and one only needs the ratio
$|z_{c2}/z_{c3}|$ to be close to 1. This is the possibility discussed in detail in section 8 and 9. 

\item[ii)] One of the right-handed neutrinos is particularly light and, in first approximation, it is only coupled to $\mu$ and
$\tau$. Thus, $M_c\sim \eta$  (small) and $z_{c1}\sim 0$. In this case
\cite{hall} the 23 subdeterminant vanishes, and one only needs the ratio
$|z_{c2}/z_{c3}|$ to be close to 1. This possibility has been especially emphasized in refs. \cite{king}. 
\end{enumerate}

In particular mechanism ii) is compatible with symmetric mass matrices. For example, one could want to preserve left-right
symmetry at the GUT scale. Then, the observed smallness of left-handed mixings for quarks would also demand small right-handed
mixings. So we now assume that $m_D$ is nearly diagonal (always in the basis where charged leptons and $M$ are diagonal)
with all its off diagonal terms proportional to some small parameter $\epsilon$. Starting from
\beq
 m_D\propto 
\left[\matrix{\epsilon^p&x\epsilon\cr x\epsilon&1    } 
\right],~~~~~ M^{-1}\propto 
\left[\matrix{r_2&0\cr 0&1    } 
\right]~~~~~. 
\label{mdM}
\eeq where $x$ is O(1), we obtain:
\beq  m_{\nu}=m^T_D M^{-1}m_D\propto 
\left[\matrix{\epsilon^{2p}r_2+x^2\epsilon^2&x\epsilon^{p+1}r_2+x\epsilon \cr x\epsilon^{p+1}r_2+x\epsilon&x^2\epsilon^2
r_2+1    } 
\right]~~~~~. 
\label{mnM}
\eeq For sufficiently small $M_2$  the terms in $r_2$ are dominant. For $p=1,2$, which we consider as typical cases, it is
sufficient that $\epsilon^2 r_2\gg1$. Assuming that this condition is satisfied, consider first the case with p=2. We have
\beq  m_{\nu}=m^T_D M^{-1}m_D\propto  x^2\epsilon^2 r_2\left[\matrix{\frac{\epsilon^2}{x^2}&\frac{\epsilon}{x}\cr
\frac{\epsilon}{x}&1    } 
\right]~~~~~. 
\label{mnp2}
\eeq This case is qualitatively similar to that described by eq. (\ref{eq3}). The determinant is naturally vanishing  (to the
extent that the terms in $r_2$ are dominant), so that the mass eigenvalues are widely split. However the mixing is nominally
small: $\sin{2\theta}$ is of order $2\epsilon/x$. It could be numerically large enough if $1/x\sim 2-3$ and $\epsilon$ is of
the order of the Cabibbo angle $\epsilon\sim 0.20-0.25$. This is what we call "stretching": the large neutrino mixing is
explained in terms of a small parameter which is not so small and can give a perhaps sufficient amount of mixing if enhanced by
a possibly large coefficient. 

A more interesting case is obtained for $p=1$, that gives:
\beq  m_{\nu}=m^T_D M^{-1}m_D\propto 
\epsilon^2 r_2\left[\matrix{1&x\cr x&x^2    } 
\right]~~~~~. 
\label{mnp1}
\eeq In this case the small parameter $\epsilon$ is completely factored out and for $x\sim 1$ the mixing is nearly maximal. The
see-saw mechanism has created large mixing from almost nothing: all relevant matrices entering the see-saw mechanism are nearly
diagonal . Clearly, the crucial factorization of the small parameter $\epsilon^2$ only arises for $p=1$. It is possible to
extend the previous model to the 3 by 3 case and to support the resulting almost diagonal textures by means of suitable flavour
symmetries \cite{afm}. In a similar class of models all Dirac mixings are small, but large mixing are introduced
via $M$ \cite{bcr}. 

Notice that we do not need a symmetric matrix $m_D$ as that given in eq. (\ref{mdM}) to get the desired effective neutrino mass
matrix.
The relevant condition is $(m_D)_{22}=(m_D)_{23}$, which can be realized in a asymmetric Dirac mass matrix. For instance
we can equally well take
\beq
m_D\propto 
\left[\matrix{\epsilon &x\epsilon\cr y &1    } 
\right]~~,
\label{asy}
\eeq
with $y\le 1$ arbitrary. For $\epsilon^2 r_2\gg1$ we end up again with $m_\nu$ given in eq. (\ref{mnp1}).

The previous models are all in trouble if the atmospheric neutrino mixing is exactly maximal or very close to maximal. There
are a few cases in the literature, not of compelling elegance, where the mixing is arranged to be exactly maximal. For
instance, in ref. \cite{wet} a discrete group generated by the elements:
\beq  R=\um
\left[
\matrix{ 0& \sqr& \sqr\cr
\sqr& 1& 1\cr
\sqr&-1&1}
\right]~~,~~~~~~~ T_n=
\left[
\matrix{ 1&0&0\cr 0&e^{2\pi i\over n}&0\cr 0&0&e^{-{2 \pi i \over n}}}
\right]~~,~~~~~(n \ge 2)
\label{wet1}
\eeq  has been suggested to constrain the leading order form of the mass matrices. Indeed, if $R$ and $T_n$ act on the
neutrinos in their flavour basis and if the required Higgs multiplets are singlets under the discrete symmetry, then, for
$n>2$, the most general invariant mass matrix for  the light neutrinos has the form:
\beq  m_\nu=m
\left[
\matrix{ 1&0&0\cr 0&0&1\cr 0&1&0}
\right]~~.
\label{wet2}
\eeq  that reproduces the case $C3$ in table I (first column), that is the degenerate case with a double maximal mixing. By
invoking an appropriate transformation property for the relevant Higgs doublet,  it is possible to obtain a hierarchical and
diagonal mass matrix  in the charged lepton sector. A suitable spontaneous breaking of the discrete symmetry is then required
to generate the correct pattern of masses and mixings beyond the leading approximation. Another example is found in ref. \cite{BHKR}.

\section{Outlook and Conclusion}

By now there are rather convincing experimental indications for neutrino oscillations.  If so, then neutrinos have non zero
masses. As a consequence, the phenomenology of neutrino masses and mixings is brought to the forefront.  This is a very
interesting subject in many respects. It is a window on the physics of GUTs in that the extreme smallness of neutrino
masses can only be explained in a natural way if lepton number is violated.  Then neutrino masses are inversely
proportional to the large scale where lepton number is violated. Also, the pattern of neutrino masses and mixings can
provide new clues on the long standing problem of quark and lepton mass matrices. The actual value of neutrino  masses is
important for cosmology as neutrinos are candidates for hot dark matter: nearly degenerate neutrinos with a common mass
around 2 eV would significantly contribute to the matter density in the universe. 

While the existence of oscillations  appears to be on a solid ground, many important experimental ambiguities remain. The
most solid indication is that  atmospheric neutrino oscillations occur with nearly maximal mixing and $\Delta m^2$ in the
$10^{-3}-10^{-2}~\eV^2$ range. It is not  clear if $\nu_{\mu}$ disappear into mainly $\nu_{\tau}$ or into some unknown
sterile neutrino kind (disappearance into $\nu_e$ is disfavoured by the data). For solar neutrinos  it is not yet clear
which of the three solutions, MSW-SA, MSW-LA and VO, is true, and the possibility also remains of different solutions if 
not all of the experimental input is correct (for example, energy independent solutions are resurrected if the
Homestake result is modified).  Finally a confirmation of the LSND alleged signal is necessary.  In our discussion here we
assumed that the LSND evidence will fade away, so that we could restrict to the 3 known kinds of neutrinos  without new
sterile species. 

The three neutrino mass eigenvalues can either be nearly degenerate or widely split.  Solutions with three nearly
degenerate neutrinos with a common mass around 2 eV, as for neutrinos of cosmological relevance, are strongly constrained
by the bound on neutrinoless double beta decay, and only nearly maximal mixing for solar neutrinos is allowed.  In any
case, because of the smallness of $\Delta m/m$, solutions of this kind are not easy to theoretically justify in a natural
way. So it looks that probably neutrinos are not cosmologically relevant. 

We argued in favour of widely split solutions for neutrino masses. Reconciling large splittings with large mixing(s)
requires some  natural mechanism to implement a vanishing determinant condition. This can be obtained in the see-saw
mechanism if one light right handed neutrino is dominant, or a suitable texture of the Dirac  matrix is imposed by an
underlying symmetry. In a GUT context, the existence of  right handed neutrinos indicates SO(10) at least as a
classification group generated by physics at $M_{Pl}$.  
The symmetry group at $M_{GUT}$ could be either (Susy) SU(5) or SO(10)  or a larger group. We have
presented a class of natural models where large right  handed mixings for quarks are transformed into large left handed
mixings for  leptons by the approximate transposition relation $m_d=m_e^T$ which is often realized in  SU(5) models. We
have shown that these models can be naturally implemented  by simple assignments of U(1) horizontal charges. 

While we favour models based on asymmetric mass matrices, (approximately) symmetric matrices  as, for example, produced in
left-right symmetric models are not excluded. We have seen that, in the see-saw mechanism, it is even possible to have
nearly maximal mixing starting from all Dirac matrices nearly diagonal in the basis where the RR Majorana matrix is
diagonal. Alternatively the large neutrino mixing could be generated by an enhancement of formally small terms. This is
because the typical small term in quark or charged lepton mass matrices is of the order of the Cabibbo angle $\lambda\sim
0.22$ which is not that small. This enhancement can be the result of a coherent addition of not so small coefficients from
diagonalizing the charged leptons and from the neutrino matrix itself. Or it could arise from a small exponent of
$m_{\mu}/m_{\tau}$ for example arising from suitable values of abelian charges.

In conclusion the fact that some neutrino mixing angles are large, while surprising at the start, was eventually found to
be well compatible, without any major change, with our picture of quark and lepton masses within GUTs. Rather it
provides us with new important clues that can become sharper when the experimental picture will be further clarified.

\vspace*{0.8cm}
\noindent
{\bf Acknowledgements}

\noindent
We would like to thank David Dooling for pointing out to us a miss--print in Table 1, in the previous version of this
review.

\vfill
\end{document}